# Kinetic study of the oxide-assisted catalyst-free synthesis of Silicon nitride nanowires

J. Farjas[a)], Chandana Rath [a,b)], A. Pinyol [c)], P. Roura[a)] and E. Bertran[c)]

a) GRMT, Department of Physics, University of Girona, Campus Montilivi, E17071 Girona, Catalonia, Spain.
b) School of Material Science and Technology, Institute of Technology, Banaras Hindu University, Varanasi, India
c) FEMAN, Departament de Física Aplicada i Optica, Universitat de Barcelona, Av. Diagonal 647, E08028 Barcelona, Catalonia, Spain.

Corresponding author e-mail: jordi.farjas@udg.es,   FAX: 0034-972-418098

**Abstract**

The synthesis of $Si_3N_4$ nanowires from the reaction of silicon nanoparticles with $N_2$ in the 1200-1440ºC temperature range is reported. The nitridation conditions are such that the reaction with nitrogen is favoured by the presence of silicon oxide in the particles and by the active oxidation of silicon without a catalyst. It is shown that the Si to $Si_3N_4$ conversion rate depends on the amount of silicon particles used in the experiments and that, in general, the reaction slows down for greater amounts. This trend is explained by particle stacking, which restricts the exchange of gases between the furnace atmosphere and the atmosphere around the inner particles. In a first stage, local oxygen partial pressure increases around the inner particles and inhibits nitridation locally. If the amount of reactant Si nanoparticles is small enough, this extrinsic effect is avoided and the intrinsic nitridation kinetics can be measured. Experiments show that intrinsic kinetics does not depend on temperature.

## 1. Introduction

$Si_3N_4$ is a wide band gap semiconductor that possesses excellent thermo-mechanical properties. $Si_3N_4$ nanowires ($Si_3N_4$-NW) also exhibit enhanced mechanical [1,2] and novel electrical and optical properties [3]. Accordingly, several synthesis techniques have been developed. The main synthesis method is carbothermal reduction and nitridation of a mixture containing silicon oxide [4-6], although other techniques are also applied: combustion synthesis [7,8], carbon-nanotube-confined chemical reaction [9,10], catalyst and catalystless reactions of silicon with nitrogen [11,12], reaction of silicon and silicon oxide with ammonia [13-15], chemical vapor deposition (CVD)[16] and reaction of liquid silicon with nitrogen [17]. The process known as oxide-assisted catalyst-free synthesis is one of the most promising because of its simplicity and high yield [14,18]. Si, $SiO_2$, SiC and metal oxide nanowires have also been grown by this method [14,19-21]. This technique is based on a CVD mechanism that involves SiO formation, via silicon active oxidation, and its reaction with nitrogen in the gas phase [14,22]. The actual mechanism is not fully understood and, as far as we know, no kinetic studies have been published hitherto. Kinetic studies on the nitridation of Si compacts have delivered a myriad of models that often relate to a particular experimental setup [23]. Such a variety of results can be attributed to the noteworthy influence of particle size, particle distribution, oxygen partial pressure and some impurities on the reaction kinetics [23,24].

In this paper we report the first kinetic results related to the oxide-assisted catalyst-free synthesis of $Si_3N_4$-NWs. Once the extrinsic effect of particle stacking was avoided, the intrinsic kinetics was measured and we discovered that kinetics was independent of temperature.

## 2. Experimental

Kinetic data of the mass evolution versus time have been obtained using a thermo-balance (Mettler Toledo model TGA850LF). The furnace is an alumina tube. Samples are kept inside an alumina crucible with a capacity of $0.9 \times 10^{-6}$ $m^3$. They are kept at a given constant temperature for one hour under a continuous flow of $N_2$ and Ar. This isothermal period is reached at 100 K/min under a continuous flow of Ar to prevent any reaction between Si and $N_2$ during heating. Although high purity Ar and N2 ($O_2$ and $H_2O$ < 5 ppm) have been used, $O_2$ from the external atmosphere reaches the furnace. The $O_2$ partial pressure is controlled by the nitrogen flow rate and measured with a mass spectrometer. The higher the $N_2$ flow, the lower the $O_2$ partial pressure is. For the reported experiments $P_{O2}=2\times10^{-3}$ atm.

Silicon nanoparticles [25] (Si-NP) grown by plasma-enhanced chemical vapor deposition have been used. Nitrogen adsorption by the BET (after Brunauer, Emmet and Teller) technique has indicated a specific surface of 56.3 m$^2$/g. The oxygen content, determined from elementary analysis and thermogravimetry, is 0.29 oxygen atoms per silicon atom ($Si_1O_{0.29}$). No catalyst is used. The structure of the reaction products has been analyzed by X-ray diffraction (XRD), high resolution transmission and scanning electron microscopy (HRTEM and SEM) and selected area electron diffraction (SAED).

## 3. Results and discussion.

*3.1 Structure of the reaction product*

The SEM micrograph of Fig. 1 shows the typical morphology of the reaction product. It is mainly constituted by nanowires with diameters of 50-350 nm and lengths ranging from 5 to 50 microns. HRTEM and SAED [22] reveal that these nanowires are α-$Si_3N_4$ single crystals. This morphology is expected when the reaction takes place in the gas phase [24]. A minor fraction of α-$Si_3N_4$ is constituted by particles, this product being less abundant when temperature is increased. In addition to α-$Si_3N_4$, minor amounts of β-$Si_3N_4$, silicon oxide and unreacted silicon have also been detected by XRD (Fig. 2).

*3.2 Dependence of the nitridation rate on the amount of sample*

Several nitridation experiments have been carried out under identical conditions of temperature (1350ºC) and oxygen partial pressure, but with a progressive decrease of the sample mass. Typical thermograms are shown in Fig. 3 where the degree of nitridation is plotted versus time. It is evident that results depend critically on the amount of silicon particles. Nitridation proceeds faster for lower masses and, in this case, the final degree of Si to $Si_3N_4$ conversion is higher.

After one hour (not shown in Fig. 3), the degree of conversion for the sample mass of 6.15 mg is 81%; however, when the mass is reduced to one half (3.00 mg), the reaction practically reaches completion. This result is confirmed, independently, by XRD characterization of the reaction product. In Fig. 2, the diffraction patterns corresponding to these two experiments are shown. The amount of unreacted silicon is clearly higher for the sample mass of 6.15 mg than it is for 3.00 mg, where it can be considered as a residual phase. If the sample mass is further reduced to 0.86 mg, the reaction is even faster and reaches completion in the very short time of 15 min (Fig. 3). Apart from this quantitative trend, note in Fig. 3 that the shape of the thermograms becomes sigmoidal when the mass is increased. This kind of kinetics has often been reported by other authors who followed different synthesis routes of $Si_3N_4$ [23, 26]. Our results clearly indicate that this apparent kinetics may be due to extrinsic parameters related to the amount of particles. In fact, for our particular experimental conditions, the sigmoidal shape disappears for sample masses lower than 1.5 mg. At this low mass, owing to the large section of the crucible that contains the sample, the Si-NPs do not cover the crucible's base completely. Therefore, the change in the kinetics is related to particle stacking. Such an explanation has been previously given by Pigeon and Varma [23] to explain the nitridation of silicon compacts (reaction-bonded silicon nitride). They argued that, at high reaction temperatures, silicon particles and reaction products sinter and modify the macroporus structure as the reaction proceeds. In particular, the inner particles react more slowly due to the difficulty with which the reaction gases reach their surface.

Although, in general terms, the interpretation of Pigeon and Varma [23] can be transferred to our results, the particular effect of the barrier to gas transport on the local conditions is very different in our case due to the actual nitridation mechanism. In their experiments, carried out at extremely low oxygen partial pressure ($<5\times10^{-8}$ atm) and in the presence of $H_2$, nitridation proceeds by direct reaction between Si and $N_2$. In contrast, at the high oxygen concentration of our experiments ($2\times10^{-3}$ atm), the reaction of SiO gas with $N_2$ is the main source of $Si_3N_4$ [22]. This essential difference in the mechanism has a pronounced effect on the reaction kinetics. For example, they always observe an initial slow stage of approximately 2 hours while, in our case, this stage is significantly short or simply does not exist (curve of 0.86 mg in Fig. 3). This initial stage is related to the decomposition of the $SiO_2$ layer (a process called 'devitrification' in ref.[23]) which acts as a barrier. In our experiments, the $SiO_2$ layer does not act as a barrier; it enhances the reaction [22]. Consequently, for a low amount of particles, nitridation begins at a high rate without any delay.

The local conditions for nitridation in our experiments can be deduced from the presence of $SiO_2$ in the reaction products. The XRD curves of Fig. 2 reveal the formation of cristobalite, and its volume fraction is higher for the experiments carried out with a higher mass. Consequently, a lower yield of $Si_3N_4$ is related to the formation of cristobalite. We can explain this relationship if we take into account the equilibrium of reaction

$$SiO_2 + Si \rightarrow 2\ SiO \qquad (1)$$

which is the first step towards nitridation [4]

$$3\ SiO + 2\ N_2 \rightarrow Si_3N4 + 3/2\ O_2 \qquad (2)$$

in our experimental conditions [22] (high oxygen partial pressure and high oxygen content in the Si-NP). In addition to restricting the $N_2$ flow, particle stacking restricts the $O_2$ removal. For the outer particles, the oxygen partial pressure ($2\times10^{-3}$ atm) is kept slightly below the threshold of passive oxidation ($7\times10^{-3}$ atm at 1400ºC [27]) while the SiO supply continues until Si exhaustion due to reaction (1) or to active oxidation. In contrast, for the inner particles, the $O_2$ resulting from decomposition of $SiO_2$ locally increases the oxygen partial pressure and, eventually, results in the formation of $SiO_2$ instead of $Si_3N_4$. However, this phenomenon does not stop nitridation. The oxygen partial pressure difference between the furnace atmosphere and the gas around the inner particles will progressively decrease its local concentration and allow the formation of SiO which will react with $N_2$ to form $Si_3N_4$. The different conditions for the inner and outer particles is nicely shown in Fig. 4, which is a cross section of the reaction product obtained from Si-NP kept at 1350ºC during 15 minutes (mass 5.29 mg). The white product, located on the upper part of the crucible, is mainly α-$Si_3N_4$, while the grayish (in fact, brownish) region is full of unreacted nanoparticles.

*3.3 Intrinsic particle-nitridation*

Once the influence of particle stacking on the reaction kinetics is established, our aim is to characterize it in the absence of this extrinsic effect. Experiments have been carried out at different temperatures in the 1250-1420ºC range with a small amount of sample (<1.5 mg). In Fig. 5, the conversion degree as a function of time has been plotted showing that the intrinsic kinetics is independent of temperature. This means that the reaction is not controlled by diffusion in the solid state, in accordance with the proposed nitridation mechanism. According to reactions (1) and (2), $N_2$ reacts with SiO. Since both species are gases, their reaction has a very low dependence on temperature.

The same behavior has also been observed by Pigeon and Varma [23]. However, their independence from temperature is limited to the first stages (below 20% conversion). It is interesting to note that they propose the volatilization of the initial silicon oxide layer as a possible explanation for this behavior. Although complementary experiments are needed, it is reasonable to propose that the initial fast stage observed in the nitridation kinetics (Fig. 5) is related to the formation of SiO gas from the initial oxygen content in the nanoparticles. After exhaustion of SiO, the kinetics slows down and depends on the rate of active oxidation.

When the series of experiments of Fig. 5 is repeated with a mass high enough to make the particles inside the crucible overlap, the result is very different (Fig. 6). Here, a pronounced dependence on temperature is apparent. In spite of that, and according to the previous discussion, we now know that this usual behavior is not intrinsic to the nitridation kinetics.

## 4. Conclusions

The intrinsic kinetics of the synthesis of $Si_3N_4$ nanowires by reaction with $N_2$ in the active-oxidation regime has been measured for the first time. Our first results indicate a very fast reaction rate during the initial five minutes, followed by a progressively slower evolution. No significant differences have been observed in the 1200-1440ºC temperature range. This independence of temperature agrees with the proposed reaction mechanism in the vapor phase between SiO and $N_2$. Complementary studies are to be performed in the near future with the aim of characterizing the observed kinetics, i.e. its dependence on: 1) the concentration of reactive gases ($N_2$ and $O_2$); b) particle size; and c) the amount of $SiO_2$ in the particles. The rate-limiting step should also be identified.

## 5. Acknowledgments

This work has been supported by the Spanish Programa Nacional de Materiales under agreement number MAT99-0569-C02. Chandana Rath wishes to acknowledge the Ministerio de Educacion y Cultura, Government of Spain her fellowship and Albert Pinyol wishes to acknowledge the European Comission for awarding a Marie Curie Fellowship under contract G1TR-CT2000-00037

# 6. References.

[1] Y. Zhang, N. Wang, R. He, Q. Zhang, J. Zhu and Y. Yan, J. Mater. Res. **15**, 1048 (2000).

[2] I.W. Chen and A. Rosenflanz, Nature **389**, 701 (1997).

[3] L. Zhang, H. Jin, W. Yang, Z. Xie, H. Miao and L. An, Appl. Phys. Lett. **86**, 061908 (2005).

[4] P.D. Ramesh and K.J. Rao, J. Mater. Res. **9**, 2330 (1994).

[5] M-J. Wang and H. Wada, J. Mat. Sci. **25**, 1690 (1990).

[6] X.C Wu, W.H. Song, W.D. Huang, M.H. Pu, B. Zhao, Y.P. Sun and J.J. Du, Mat. Res. Bull. **36**, 847 (2001).

[7] YG. Cao, CC. Ge, ZJ. Zhou and J.T. Li, J. Mat. Res. **14**, 876 (1999).

[8] H. Chen, Y. Cao, X. Xiang, J. Li and C. Ge, J. Alloy. Compd. **325**, L1 (2001).

[9] W. Han, S. Fan, Q. Li, B. Gu, X. Zhang and D. Yu, Appl. Phys. Lett. **71**, 2271 (1997).

[10] W. Han, S. Fan, Q. Li and Y. Hu, Science **277**, 1287 (1997).

[11] P.S. Gopalakrishnan and P.S. Lakshminarasimham, J. Mater. Sci. Lett. **12**, 1422 (1993).

[12] Y.G Cao, H. Chen, J.T. Li, C.C. Ge, S.Y. Tang, J.X. Tang and X. Chen, J. Cryst. Growth **234**, 9 (2002).

[13] H.Y. Kim, J. Park, H. Yang, Chem. Phys. Lett. **372**, 269 (2003).

[14] Y. Zhang, N. Wang, R. He, J. Liu, X. Zhang, J. Zhu, J. Cryst. Growth **233**, 803 (2001).

[15] LW Yin, Y. Bando, YC Zhu and YB Li, Appl. Phys. Lett. **83**, 3584 (2003).

[16] S. Motojima, T. Yamana, T. Araki and H. Iwanaga, J. Electrochem. Soc. **142**, 3141 (1995).

[17] Y. Inomata, T. Yamane, J. Cryst. Growth **21**, 317 (1974).

[18] G.Z. Ran, L.P. You, L. Dai, Y.L. Liu, Y. Lv, X.S. Chen and G.G. Qin, Chem. Phys. Lett. **384**, 94 (2004).

[19] D. D. D. Ma, C.S. Lee, F.C.K. Au, S.Y. Tong and S.T. Lee, Science **299**, 1874 (2003).

[20] Y. Zhang, N. Wang, S. Gao, R. He, S. Miao, J. Liu, J. Zhu and X. Zhang, Chem. Mater. **14**, 3564 (2002).

[21] H.Y. Dang, J. Wang and S.S Fan, Nanotechnology, **14**, 738 (2003).

[22] Farjas J., Rath C., Pinyol A., Roura P. and Bertran E., Appl. Phys. Lett., **87**, 192114 (2005).

[23] R.G. Pigeon and A. Varma, J. Mat. Sci. **28**, 2999 (1993).

[24] A. J. Moulson, J. Mat. Sci. **14**, 1017 (1979).

[25] J. Costa, G. Sardin, J. Campmany, and E. Bertran, Vacuum **45**, 1115 (1994).

[26] B.W. Sheldom, J. Szekely and J.S. Haggerty, J. Am. Ceram. Soc. **75**, 677 (1992).

[27] R.S Parikh, A. Ligtfoot, J.S. Haggerty, B.W Sheldon, J. Am. Ceram. Soc. **82**, 2626 (1999).

**Figure captions**

**Fig. 1**. SEM micrograph obtained from Si-NP kept at 1350ºC during one hour under a continuous flow of $N_2$ and Ar. The oxygen partial pressure is $2\times10^{-3}$ atm. The sample mass is 3.00 mg. The first stages of transformation are plotted in Fig. 3.

**Fig 2.** XRD pattern obtained from Si-NP kept at 1350ºC during one hour under a continuous flow of N2 and Ar for different sample quantities. The first stages of transformation are plotted in Fig. 3

**Fig 3**. Effect of particle stacking on the nitridation kinetics. Thermograms obtained from Si-NP kept at 1350ºC under a continuous flow of $N_2$ and Ar for different sample quantities. The degree of conversion is directly calculated from the measured mass gain.

**Fig 4**. Cross section of the reaction product obtained from Si-NP kept at 1350ºC during 15 minutes. Sample mass 5.29 mg.

**Fig 5**. Thermograms obtained from Si-NP at different temperatures under a continuous flow of $N_2$ and Ar. The sample amounts are small enough to prevent particle overlapping. Under these conditions, the kinetics can be considered as intrinsic to the chemical reaction. Note its independence from temperature.

**Fig. 6**. Thermograms obtained under the same conditions as those of Fig. 5 but with a higher mass. The stacking of particles, in this case, has a pronounced extrinsic influence on the nitridation kinetics leading to an apparent dependence on temperature.

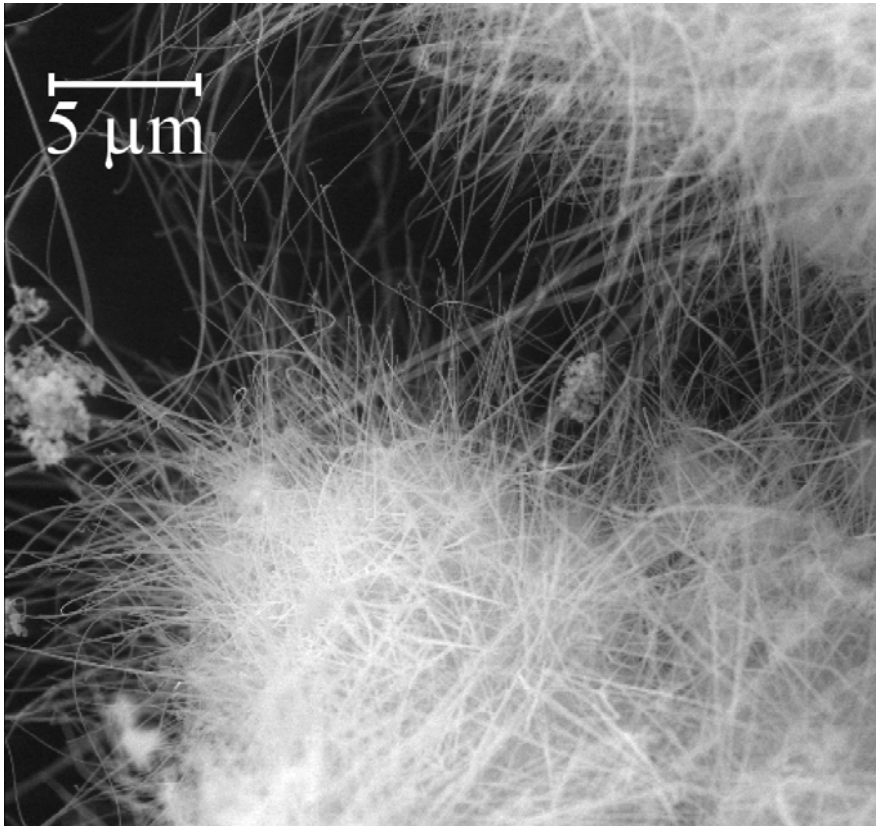

Fig. 1.

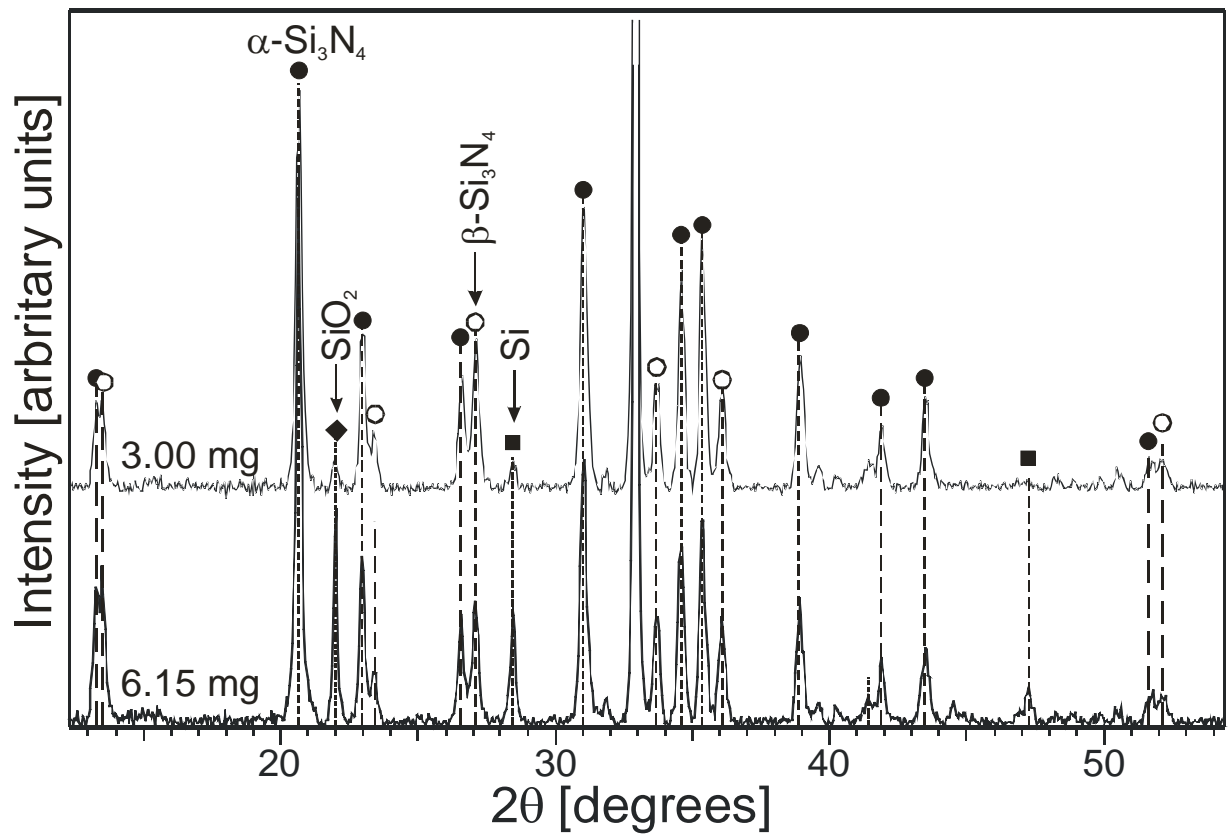

Fig. 2.

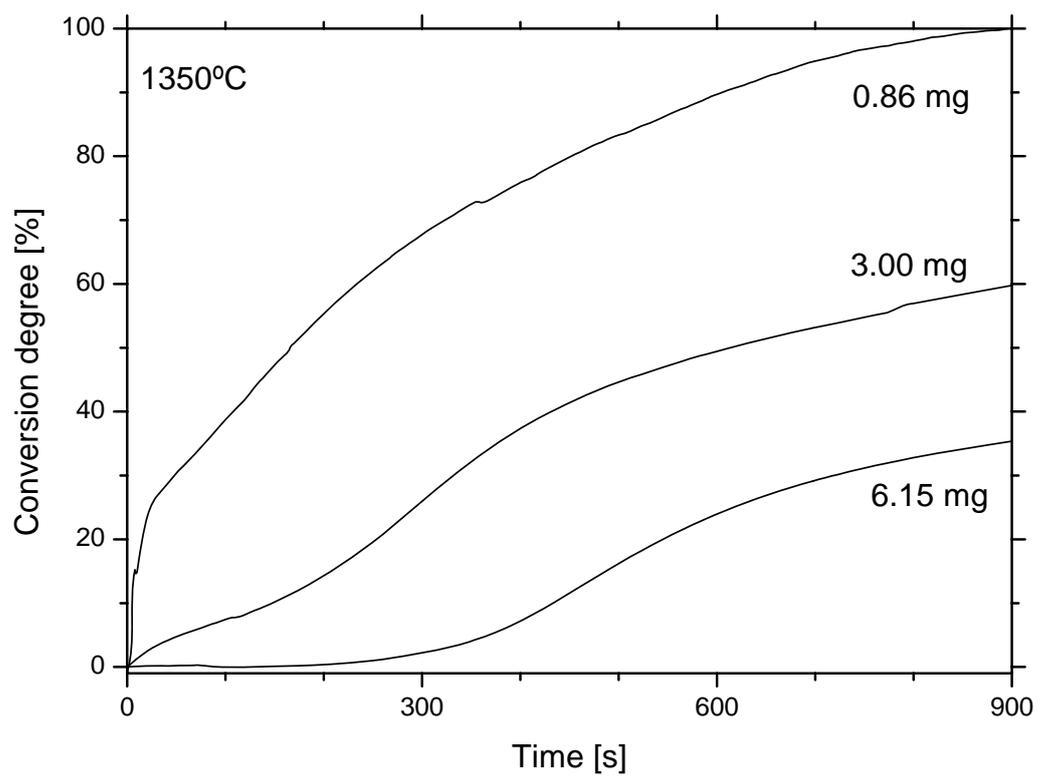

Fig 3.

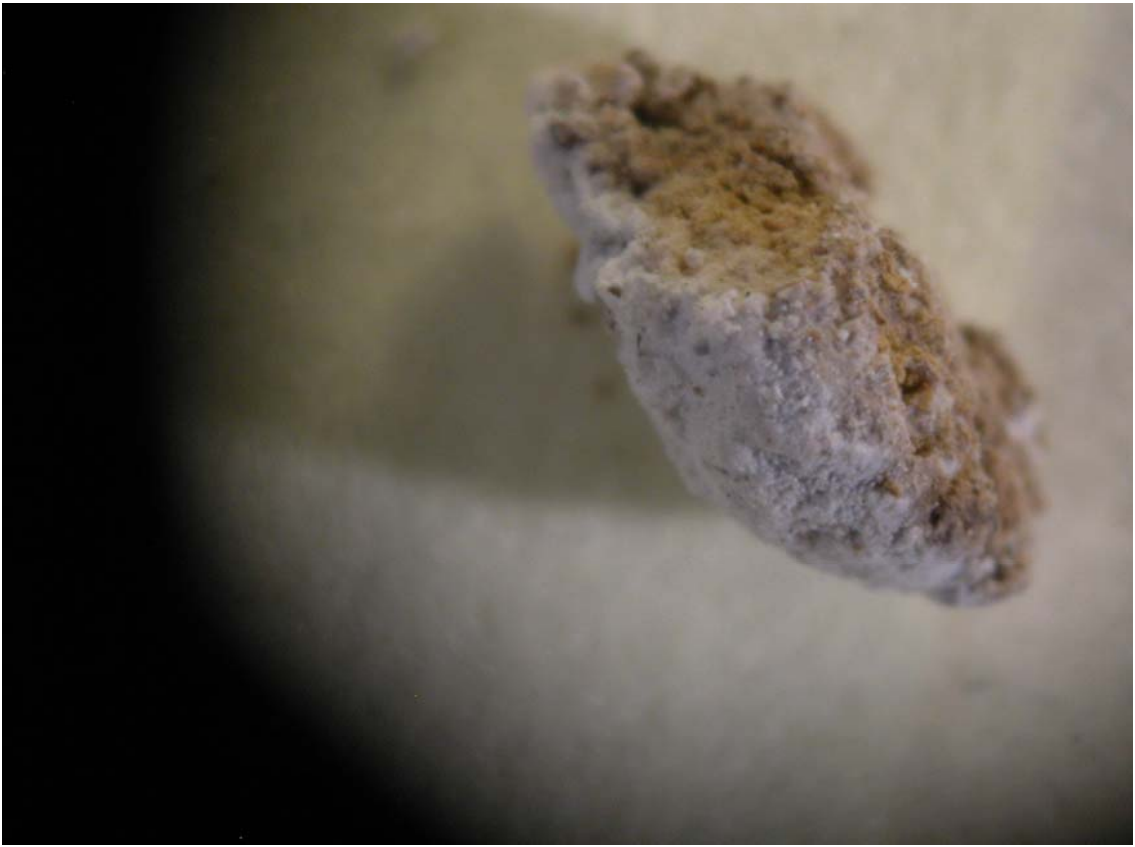

Fig. 4.

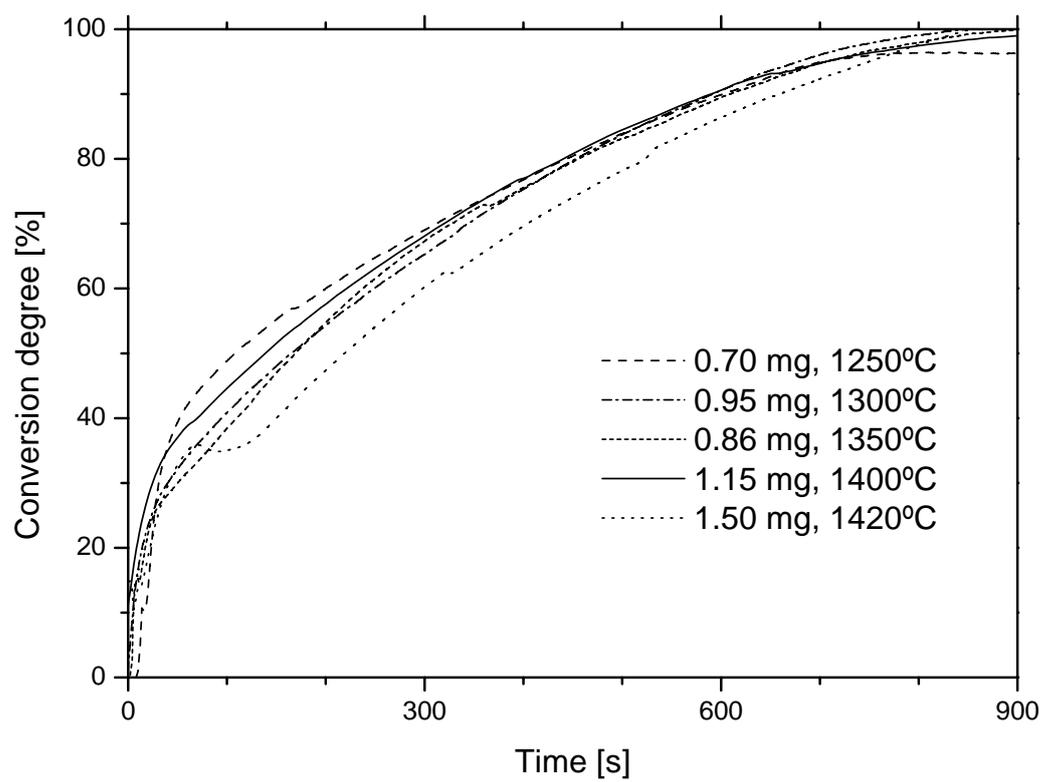

Fig. 5.

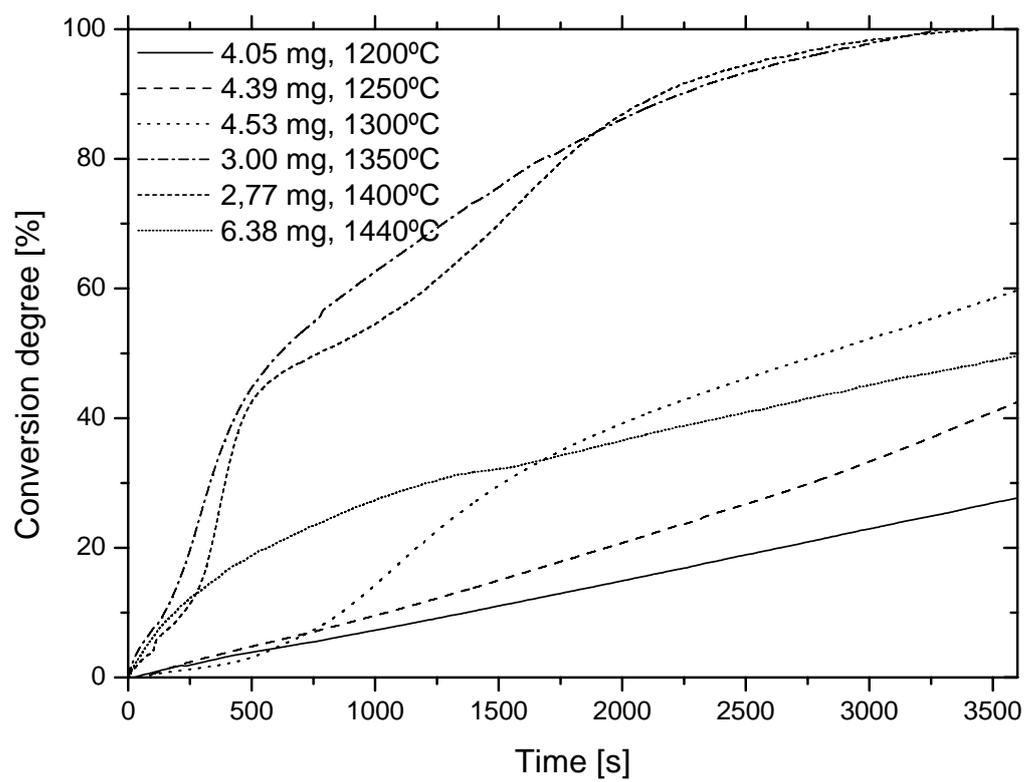

Fig. 6.